\documentclass{llncs}
\usepackage[table,xcdraw]{xcolor}
\usepackage{makeidx}  
\usepackage[english]{babel}
\usepackage[utf8]{inputenc}
\usepackage{amsmath}
\usepackage{graphicx}
\usepackage[colorinlistoftodos]{todonotes}
\usepackage{amssymb}
\usepackage{color}
\usepackage{graphicx}
\usepackage{amsmath}
\usepackage{amssymb}
\usepackage{subfigure}
\usepackage{latexsym}
\usepackage{color}
\definecolor{blue}{rgb}{0,0,1}

\usepackage[algo2e,ruled,vlined]{algorithm2e}
\begin{document}
\newtheorem{conjeture}[theorem]{Conjeture}
\mainmatter              
\title{Separating Topological Noise from Features using Persistent Entropy}
\titlerunning{Separating Topological Noise from Features}  
%
\author{Nieves Atienza\inst{1} \and Rocio Gonzalez-Diaz\inst{1}
\and Matteo Rucco\inst{2}}
\authorrunning{N. Aitenza, R. Gonzalez-Diaz and M. Rucco} 
%
%
\institute{Applied Math Department, School of Computer Engineering, University of Seville, Seville, Spain,\\
\email{$\{$natienza,rogodi$\}$@us.es},\\
\and
Univ. of Camerino, School of Science and Technology, Computer Science Division,
Camerino, IT\\
\email{matteo.rucco@unicam.it}}

\maketitle              

\begin{abstract}
Persistent homology appears as a fundamental tool in Topological Data Analysis.
It  studies the evolution of $k-$dimensional holes along a sequence of simplicial complexes (i.e. a filtration).
The set of intervals representing birth and death times of 
$k-$dimensional holes along such sequence
 is called the persistence barcode.
$k-$dimensional holes with short lifetimes are informally considered to be “topological noise”, and those with a long lifetime are considered to be “topological feature” associated to the given data (i.e. the filtration).
In this paper, we derive a simple method for separating topological noise from topological features using a novel measure for comparing persistence barcodes called {\it persistent entropy}.

\end{abstract}
\section{Introduction}
 Persistent homology  study the evolution of $k-$dimensional holes along a sequence of simplicial complexes.
Persistence barcode is the collection of intervals representing birth and death times of 
$k-$dimensional holes along such sequence.
In persistence barcode,  $k-$dimensional holes with short lifetimes are informally considered to be “topological noise”, and those with a long lifetime are  “topological feature” of the given data. 

In \cite{confidence} a methodology is presented
for deriving confidence sets for persistence diagrams. They derived a simple method for separating topological noise from topological features. 
The authors focused on simple, synthetic examples as proof
of concept. They presented several methods for separating noise
from features in persistent homology. 
The
methods has a simple visualization: one only needs to add a band around the
diagonal of the persistence diagram. Points in the band are consistent with
being noise.
The first three methods are based on the
distance function to the data. 
They start with a sample from a distribution $\mathbb{P}$ supported on a topological space $\mathfrak{C}$. The bottleneck distance is used as a metric on the space of persistence diagrams. 
The last method uses density estimation. The advantage of the former is
that it is more directly connected to the raw data. The advantage of the
latter is that it is less fragile; that is, it is more robust to noise and outliers.

 Persistent entropy (which is the Shannon entropy of the persistence barcode) is a tool formally  defined in  \cite{rucco2015characterisation} and used for measure similarities between two persistence barcodes.
A precursor of this denition was given in \cite{chintakunta2015entropy} to measure how dierent bars of the barcode are in length.

In this paper, we use the difference of persistent entropy  for measure similarities between two persistent barcodes, in order to derive such confidence sets.
More concretely, we derive a simple method for separating topological noise from topological features of a given persistence barcode obtained from a 
given filtration (ie., a sequence of simplicial complexes) using the mentioned persistent entropy measurement.


\section{Related work}\label{matteo}

\begin{figure}[t!]
	\centering
		\includegraphics[width=10cm]{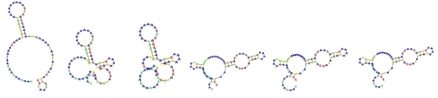}
	\caption{From left to right: RNA secondary suboptimal structures within different bacteria.}
\label{fig_RNA}
\end{figure}

Persistent homology based techniques are nowadays widely used for analyzing data especially if the data under observation are produced by an essentially nonlinear process~\cite{nanda2014simplicial}. Persistent homology is the proper tool for pinpointing out the shape of the object underlying a dataset. This can appear only a geometrical issue but, indeed, it is strongly related to the functional properties of the systems that generated the data. Let's take a look at Fig.~\ref{fig_RNA}, it represents a collection of RNA secondary suboptimal structure within different bacteria. All the shapes are characterized by several circular substructures, each of them is obtained by  linking different amminoacids. Each substructure encodes functional properties of the bacteria. Persistent homology properly deals with this classification. For the love of preciseness, Mamuye et al. \cite{Adane}, used Vietoris-Rips complexes and persistent homology for certifying that there are different species but characterized with the same RNA suboptimal secondary structure, thus these species are functionally equivalent. In~\cite{rucco2015topological}, the authors proposed a new methodology based on information theory and persistent homology for classifying real length noisy signals produced by small DC motors. They introduced an innovative approach based on ``auto mutual information'' and the ``CAO's method'' for providing the time delay embedding of signals. 
The time delay embedding transforms the signal into a point cloud data in $\mathbb{R}^d$ that is completed to the Vietoris-Rips complexes and thus analyzed by persistent homology. 
The authors classified the signal in two classes, respectively ``properly working'' and ``broken''.
However, Vietoris-Rips based analysis suffers of the selection of the parameter $\epsilon$: for different $\epsilon$ different topological features can be observed. There is the request to differentiate what are meaningful topological features from what are not, i.e. topological noise. Actually, as far as we know, some statistics have been provided and they are known as `` persistence landscape''~\cite{bubenik2015statistical}. A landscape is a piecewise linear function that basically maps a point within a persistent diagram (or barcode) to a point in which the $x-$coordinate is the average parameter value over which the feature exists, and the $y-$coordinate is the half-life of the feature. Landscape is a powerful tool for statistically assessing the global shape of the data over different  $\epsilon$, or in word it helps to see which topological features persistent for different $\epsilon$. Nevertheless, landscape can not be used for differentiating which are topological features from which are not and we claim that our method, explained in Section~\ref{method}, fills this gap.


\section{Background}
\label{Topology}

This section provides a short recapitulation of the basic concepts 
needed as a basis for the presented method for separating topological noise from features.

A topological space is a powerful mathematical concept for describing the connectivity of a space. Informally, a topological space is a set of points each of them equipped with the notion of neighboring. 
One way to represent a topological space embedded in $\mathbb{R}^n$ is by decomposing it into {\it simple} pieces such that their common intersections are lower-dimensional pieces of the same kind. In this paper, we use simplicial complexes as the data structure to represent topological spaces.
A {\it simplicial complex} $K$ is composed by a set $K_0$ of $0-$simplices (also called vertices,  that can be thought as points in $\mathbb{R}^n$);
and, for each $k\geq 1$, a set $K_k$ of $k-$simplices $\sigma=\{v_0, v_1, \dots, v_k\}$, where $v_i \in V$; satisfying that:
\begin{itemize}
\item each $k-$simplex has $k+1$ faces obtained removing one of its vertices;
\item if a simplex $\sigma$ is in $K$, then all faces of $\sigma$ must be in $K$.
\end{itemize}
The underlying space of $K$ is the union of its simplices.
 Notice that a $1-$simplex is  an edge, a $2-$simplex  a filled triangular face and a $3-$simplex  a filled tetrahedron. 
We only consider finite simplicial complexes with finite dimension, i.e., there exists an integer $n$ (called the dimension of $K$) such that for $k>n$, $K_k=\emptyset$ and for $0\leq k\leq n$, $K_k$ is a finite set. 
See \cite{hatcher2002algebraic} and \cite{munkres1984elements} for an introduction to algebraic topology.

\begin{figure}[t!]
	\centering
		\includegraphics[width=6cm]{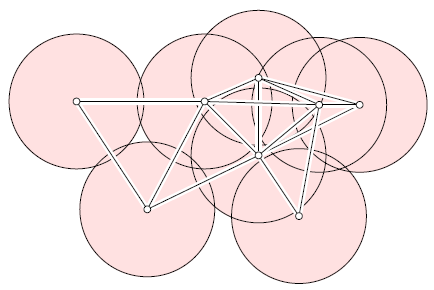}
	\caption{\cite[p. 72]{edelsbrunner2010computational} Nine points with pairwise intersections among the disks indicated by
straight edges connecting their centers, for a fixed time $\epsilon$. The 
Čech complex $\mbox{\it Č}_{\epsilon}(V)$ fills nine of the ten possible
triangles as well as the two tetrahedra. The Vietoris-Rips complex
$VR_{\epsilon}(V)$
fills the ten 
triangles and the two tetrahedra. }
	\label{figure_cech}
\end{figure}

Two classical examples of simplicial complexes are the Čech complexes and the Vietoris-Rips complexes (see \cite[Chapter III]{edelsbrunner2010computational}). 
Let $V$ be a finite set of
points in $\mathbb{R}^n$.  The {\it Čech complex} of $V$ and $r$ denoted by {\it Č}$_{r}(V)$ is  the simplicial complex whose simplices are formed as follows. For each subset $S$ of points in $V$, form   a  closed ball of radius $r/2$ around each point in $S$, and include $S$ as a simplex of {\it Č}$_{r}(V)$  if there is a common point contained in all of the balls in $S$. This structure satisfies the definition of a simplicial complex.
The {\it Vietoris-Rips complex} denoted as $VR_{r}(V)$ is essentially the same as the Čech complex, except instead of 
Instead of checking if there is a common point contained in the  intersection of  the $(r/2)-$ball around $v$ for all $v$ in $S$, we may
just check pairs
adding $S$ as a simplex of {\it Č}$_{r}(V)$ if   all the balls have pairwise intersections.  We have 
 $\mbox{\it Č}_{r}(V) \subseteq VR_r(V)\subseteq \mbox{\it Č}_{\sqrt{2} r}(V)$. See Fig.\ref{figure_cech}.

Homology is an algebraic machinery used for describing topological spaces. The $k-$Betti number $\beta_k$ represents the rank of the $k-$dimensional homology group of a given simplicial complex $K$.
Informally, 
$\beta_0$ is the number of connected components, $\beta_1$  counts the number of
loops in  $\mathbb{R}^2$ or tunnels in  $\mathbb{R}^3$ and $\beta_2$ can be thought as the number of voids.
Persistent homology  is a method for computing $k-$dimensional holes (i.e., the rank of the $k-$dimensional homology group) of $K$, at different spatial resolutions. 
If a hole is  detected over a wide range of different spatial resolutions then it is likely to represent true features of the underlying space, rather than artifacts of sampling, noise, or particular choice of parameters.
For a more formal description we refer to~\cite{edelsbrunner2010computational}.  In order to compute persistent homology, we need a distance function on the underlying space. This can be obtained constructing a filtration of the simplicial complex  which  is a collection of simplicial complexes
$\{K(t)| t \in \mathbb{R}\}$ of $K$ such that $K(t) \subset K(s)$ for $t< s$ and there exists $t_{\max}\in \mathbb{R}$ such that $K_{t_{\max}}=K$. The filtration time (or filter value) of a simplex $\sigma \in K$ is the smallest $t$ such that $\sigma \in K(t)$. 

 Given a  finite set of
points $V$ in $\mathbb{R}^n$, let $t_{max}$ be the diameter of $V$ (that is, $t_{\max}=max{d(x,y)|x,y\in V}$, then
 $\{\mbox{\it Č}_{t}(V)| t \in \mathbb{R}\}$ and $\{VR_{t}(V)| t \in \mathbb{R}\}$  are, respectively, the Čech  and the  Vietoris-Rips filtration.
Observe that if $V$ has $d$ points, then  $\mbox{\it Č}_{t_{\max}}(V)$ is equal to $VR_{t_{\max}}(V)$ and contains exactly the $(d-1)-$simplex defined by $V$ together with all its faces.

{\it Persistent homology} describes how the homology of a given simplicial complex $K$ changes along filtration. A $k-$dimensional Betti interval, with endpoints $[t_{start}, t_{end}),$ corresponds to a $k-$dimensional hole that appears at filtration time $t_{start}$ and remains until  filtration  time $t_{end}$. 
The set of intervals representing birth and death times of homology classes is called the {\it persistence barcode} associated to the corresponding filtration.

 Observe that,  in the case of the Čech  and  Vietoris-Rips filtrations, there is only one infinite interval corresponding to the connected component that appears at $t=0$ and dies at $t=d/2$ being $d$ the diameter of $V$, since at time  $t=d/2$  the balls around all the points of $V$ have pairwise intersections and then no new simplex or hole is created later. In our case, the interval associated to the connected component that survives until the end is considered to be $[0 , d/2)$.


\section{Persistent Entropy}\label{entropy}

In order to measure how much the construction of a filtered simplicial complex is ordered, a new entropy measure, the so-called \textit{persistent entropy}, were defined in~\cite{rucco2015characterisation}. A precursor of this definition was given in~\cite{chintakunta2015entropy} to measure how different bars of the barcode are in length. In \cite{signal},  persistent entropy is used  for addressing the comparison between discrete
piecewise linear functions.

Given a  Čech  or  Vietoris-Rips filtrations   $F=\{K(t) |  t\in \mathbb{R}\}$, and the corresponding persistence barcode $B = \{[a_i , b_i) : 1\leq i\leq n\}$,
let $L=\{\ell_i=b_i-a_i : 1\leq i\leq n\}$. The \textit{persistent entropy} $H$ of the filtration $F$ is:
$$H_L=-\sum_{i=1}^n \frac{\ell_i}{S_L}  log \frac{\ell_i}{S_L},\qquad \mbox{ being  }
S_L=\sum_{i\in I}\ell_i.$$
Note that the maximum persistent entropy would correspond to the situation in which all the intervals in the barcode are of equal length. Conversely, the value of the persistent entropy decreases as more intervals of different length are present. 
More concretely,  if $B$ has $n$ intervals,  the possible values of the persistent entropy $H_L$ associated with the barcode $B$ lie in the interval $[0,\log(n)]$. 

The following result supports the idea that persistence  entropy can differentiate long from short intervals as we will see in the next section.
\begin{theorem}\label{theorem}
For a fixed integer $i$, $1\leq i\leq n$, 
let $L_i=\{\ell_i,\dots\ell_n\}$,  $S_i=\sum_{j=i}^{n}\ell_i$ and let $H_{i}$ be the persistent entropy associated to $L_i$. Let  
$$L'(i)=\{\ell'_1,\dots,\ell'_{i-1},\ell_{i},\dots,\ell_n\},\quad\mbox{ where } 
 \ell'_j=S_{i}/e^{H_{i}},\quad\mbox{ for }1\leq j\leq i-1.$$
  Then 
 $H_L\leq H_{L'(i)}$.
 \end{theorem}

\proof
Let us prove that $H_{L'}$ is the maximum of all the possible persistent entropies associated to lists of intervals with $n$ elements, such that the last $i$ elements of any of such lists is $\{\ell_{i},\dots,\ell_n\}$. 
Let $M= \{x_1,\dots,x_{i-1},\ell_{i},\dots,\ell_n\}$ (where $x_j>0$ for 
$1\leq j\leq i-1$) be any of such lists.
 Let  $S_x=\sum_{j=1}^{i-1}x_i$. 
 Then,  the persistent entropy associated to $M$ is:
\begin{eqnarray*}
H_M
&=&\sum_{j=1}^{i-1}
\frac{x_j}{S_x+S_i}\log\left(\frac{x_j}{S_x+S_i} \right)
+\sum_{j=i}^n\frac{\ell_j}{S_x+S_i}\log\left(\frac{\ell_j}{S_x+S_i} \right).
\end{eqnarray*}
In order to find out the maximum of $H_M$ with respect to the unknown variables 
$x_k$, $1\leq k\leq i-1$, we compute the partial derivative of $H_M$ 
with respect to those variables:
\begin{eqnarray*}
\frac{\partial H_M}{\partial x_k}
=\frac{1}{(S_x+S_i)^2} &&\left(-S_iH_i+
S_i\log\left(\frac{S_i}{x_k}\right)
+
\sum_{j\neq k}x_j\log\left(\frac{x_j}{x_k}\right)\right).
\end{eqnarray*}
Then, $\{x_k=\frac{S_i}{e^{H_ i}}: 1 \leq k\leq i-1\}$ is the solution of the system $\{\frac{\partial H_M}{\partial x_k}=0: 1\leq k \leq i-1\}$.
\qed

\section{Separating topological features from topological noise}
\label{method}

Let us start with a sample $V$ from a distribution $\mathbb{P}$ supported on
a topological  space $\mathfrak{C}$. 
Suppose the  Vietoris-Rips filtration $F$ is computed from $V$, and the persistence barcodes $B$ is computed from $F$.
The following are the steps of our proposed method, based on persistent entropy, to separate topological noise from topological features in the persistence barcode $B$, estimating, in this way, the topology of $\mathfrak{C}$.

\begin{itemize}
\item[1.] Order the intervals in $B$ by decreasing length. Then 
$L=\{\ell_i=b_i-a_i: 1\leq i\leq n\}$ satisfies that $\ell_i\leq \ell_j$ for $i<j$;
\item[2.] Compute the persistent entropy $H_L$ of $B$. Denote $H_{L'(0)}:=H_L$.
\item[2.] From $i=1$ to $i=n$,
\begin{itemize}
\item[a.] Compute the persistent entropy $H_{L'(i)}$ for 
$L'(i)=\{\ell'_1,\dots,\ell'_{i-1},\ell_i,\dots,$ $\ell_n\}$, being
$\ell'_k=\frac{S_i}{e^{H_i}}$ for $1\leq k\leq i-1$ as in Th. \ref{theorem}.
\item[b.] Compute 
$H_{rel(i)}=(H_{L'(i)}-H_{L'(i-1)})/(\log(n)-H_L)$.
\item[c.] If $H_{rel(i)}>\frac{i-1}{n}$, then the associated interval $[a_i,b_i)$ represents a topological feature. Otherwise, the interval $[a_i,b_i)$ represents  noise. 
\end{itemize}
\end{itemize}

In step 3.a, the measure can be considered as the influence of the current interval in the  initial  persistent entropy $H_L$. 
It is in order to appreciate this influence, why we divide $H_{L'(i)}-H_{L'(i-1)}$ by the difference of the possible maximal entropy (which is $\log(n)$) and $H_L$. We compare the resulting $H_{rel}(i)$ with $\frac{i-1}{n}$ since  $H_{rel(i)}$ is affected by the total number of intervals and the number of intervals we  replace.
Finally, observe that $H_{L'(0)}=H_L$, 
 $H_{L'(i)}<H_{L'(j)}$ for $0\leq i<j\leq n$  and $H_{L'(n)}= \log(n)$
by Th. \ref{theorem}.

\begin{figure}[t!]
	\centering
		\includegraphics[width=8cm]{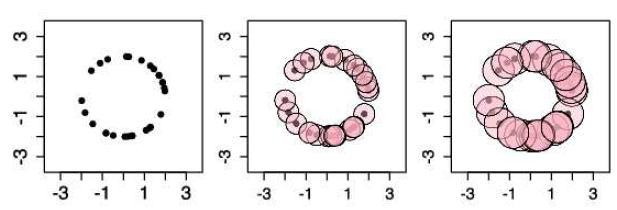}
	\caption{Left: $30$ data points sampled from a circle of radius $2$. Middle: Balls of radius $0.5$ centered at the sample points. 
Right: Balls of radius $0.8$ centered at the sample points.}
	\label{circle}
\end{figure} 

We have applied our method to two different scenarios.
First, let us take $30$ data points sampled from a circle of radius $2$ (see Fig. \ref{circle}.Left). This example has been taken from paper \cite{confidence}. For t=0.5, 
The Vietoris-Rips complex for $t=0.5$ can be deduced from the picture shown in Fig. 
\ref{circle}.Middle) which consists of 
 two connected
components and zero loops. 
Looking at the Vietoris-Rips complex for $t=0.8$ (see Fig. \ref{circle}.Right), we assist at the birth and death of topological features: at $t= 0.8$, one of the
connected components dies (is merged with the other one), and a loop appears; this loop will die at $t= 2$, when the union of the pink balls representing the distance function becomes simply connected. 

\begin{table}[]
\centering
\caption{Results of our method applied to the Vietoris-Rips filtration obtained from  $30$ data points sampled from a circle of radius $2$ (see Fig. \ref{circle}.Left). }
\label{figure_4}
\begin{tabular}{|c|c|c|c|c|c|c|c|}
\hline
 $i$ & $\ell_i$ & $\frac{\ell_i}{L}$  & $\ell'_i$ & $\frac{\ell'_i}{L'(i)}$ & $\frac{H_{L'(i)}}{\log(n)}$ 
 &$\frac{H_{L'(i)}-H_{L'(i-1)}}{\log(n)-H_L}$ & Feature \\
 \hline\hline
 1  & 2.  & 0.233645  & 0.38116  & 0.0372913 & 0.967011 
 & 0.693554 & \text{yes} \\ \hline
 2  & 1.2 & 0.0934579 & 0.325036 & 0.0349874 & 0.985761 
 & 0.174173 & \text{yes} \\ \hline
 3  & 0.7 & 0.0545171 & 0.30378  & 0.0343202 & 0.991422 
 & 0.052589 & \text{no} \\ \hline
 4  & 0.45& 0.0350467 & 0.296692 & 0.0341938 & 0.992506 
 & 0.0100741 & \text{no} \\ \hline
 5  & 0.45& 0.0350467 & 0.288896 & 0.03405   & 0.993746 
 & 0.0115148 & \text{no} \\ \hline
 \text{...} &  \text{...} & \text{...} & \text{...} & \text{...} & \text{...} & \text{...} & \text{...} \\ \hline
\end{tabular}
\end{table}

\begin{figure}[t!]
	\centering
		\includegraphics[width=10cm]{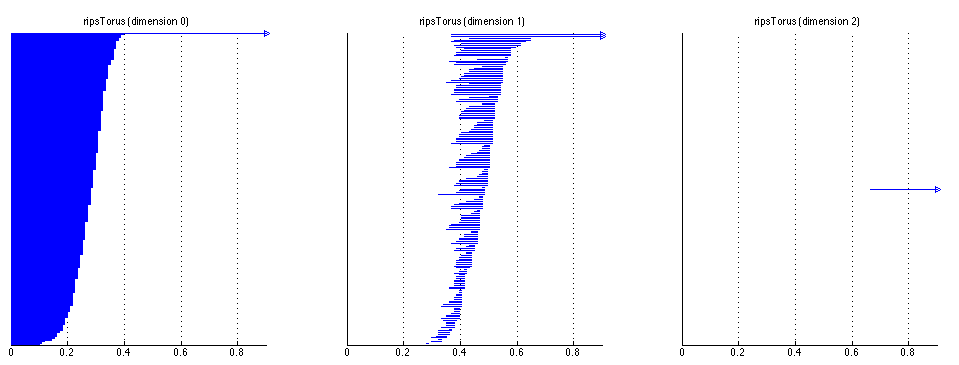}
	\caption{Barcodes (separated by dimension) computed from the Vietoris-Rips filtration associated to a point cloud lying on a 3D torus. Left: lifetimes of connected components. Middle: lifetimes of tunnels. 
Right: lifetimes of voids. }
	\label{figure_torus}
\end{figure}

  Consider now a set $V$ of $400$ points sampled from  a 3D torus.
The barcodes (separated by dimension) computed from the Vietoris-Rips filtration associated to $V$ are showed in Fig. \ref{figure_torus}.

\begin{table}[]
\centering
\caption{Application of our method to the Vietoris-Rips filtration obtained from$400$ points sampled from  a 3D torus.}
\label{figure3}
\begin{tabular}{|c|c|c|c|c|c|c|c|}
\hline
 $i$ & $\ell_i$ & $\frac{\ell_i}{L}$  & $\ell'_i$ & $\frac{\ell'_i}{L'(i)}$ & $\frac{H_{L'(i)}}{\log(n)}$ 
 &$\frac{H_{L'(i)}-H_{L'(i-1)}}{\log(n)-H_L}$ & Feature \\
 \hline\hline
 1  & 1.9      & 0.0145219  & 0.268369 & 0.00207708 & 0.971259 
 & 0.0799069  & \text{yes} \\ \hline
 2  & 1.531    & 0.0117016  & 0.262812 & 0.00205432 & 0.972992 
 & 0.0554616  & \text{yes} \\ \hline
 3  & 1.531    & 0.0117016  & 0.257239 & 0.00203115 & 0.974775 
 & 0.0570812  & \text{yes} \\ \hline
 4  & 1.234    & 0.00943158 & 0.253276 & 0.00201566 & 0.975978 
 & 0.0385369  & \text{yes} \\ \hline
 5  & 0.396    & 0.00302667 & 0.252916 & 0.00201511 & 0.976021
 & 0.00137745 & \text{no} \\ \hline
 \text{...}    &  \text{...} & \text{...} & \text{...} & \text{...} & \text{...} & \text{...} & \text{...} \\ \hline
\end{tabular}
\end{table}

\section{Conclusions and future works}\label{conclusion}
In this paper, we have derived a  method for separating topological noise from topological features using the Shannon entropy of the persistence barcode. We have prove that the method is consistent by proving that in  step $i$ of the method we replace $i$ intervals by the same number of interval but with the length that maximizes the entropy. This way we ``neutralize" the effect of such $i$ intervals and by computing the difference of the entropies obtained in step $i-1$ and step $i$, we can deduce if the interval at position $i$ is a topological feature or not. 
As a future work, we plan to apply our method separately to each $k-$persistence barcode (i.e., the barcode corresponding to the lifetimes of $k-$dimensional holes).

\bibliographystyle{elsarticle-num} 
\bibliography{typeinst}

\end{document}